\documentclass[aps,prb,reprint,superscriptaddress,amsmath,amssymb]{revtex4-2}

\usepackage{graphicx}
\usepackage{dcolumn}
\usepackage{bm}
\usepackage{xcolor} 
\usepackage{comment}
\usepackage{soul}

\newcommand{\parent}{EuAl$_4$}
\newcommand{\doped}{Eu(Al$_{0.4}$Ga$_{0.6}$)$_4$}

\newcommand{\series}{Eu(Al$_{1-x}$Ga$_{x}$)$_4$}

\newcommand{\magHelixk}{\textbf{q}$_k^{Helix}$}
\newcommand{\magHelixh}{\textbf{q}$_h^{Helix}$}
\newcommand{\magSDWh}{\textbf{q}$_h^{SDW}$}
\newcommand{\magSDWk}{\textbf{q}$_k^{SDW}$}

\begin{document}

\title{Competing collinear and non-collinear spin textures imaged by spatially-resolved REXS in \doped}


\author{Fellipe B. Carneiro}
\email{fellipe.carneiro@diamond.ac.uk}
\affiliation{Diamond Light Source, Harwell Science and Innovation
Campus, Didcot, Oxfordshire, UK}

\author{Z{\'e}t{\'e}ny Bacs{\'o}}
\affiliation{Department of Physics, Durham University, South Road, Durham DH1 3LE, United Kingdom}
\affiliation{Department of Physics and Astronomy, University College London,
Gower Street, London, WC1E 6BT United Kingdom}
\affiliation{Diamond Light Source, Harwell Science and Innovation
Campus, Didcot, Oxfordshire, UK}
\author{Kevin Allen}
\affiliation{Department of Physics and Astronomy and Rice Center for Quantum Materials (RCQM), Rice University, Houston, 77005, TX, USA}
\author{Aly H. Abdeldaim}
\author{Rebecca Scatena}
\affiliation{Diamond Light Source, Harwell Science and Innovation
Campus, Didcot, Oxfordshire, UK} 
\author{Jaime M. Moya}
\affiliation{Department of Chemistry, Princeton University, Princeton, NJ, USA}
\author{Roger D. Johnson}
\affiliation{Department of Physics, Durham University, South Road, Durham DH1 3LE, United Kingdom}
\affiliation{Department of Physics and Astronomy, University College London,
Gower Street, London, WC1E 6BT United Kingdom}
\affiliation{London Centre for Nanotechnology, University College London,
Gordon Street, London WC1H 0AH, United Kingdom}
\author{Emilia Morosan}
\affiliation{Department of Physics and Astronomy and Rice Center for Quantum Materials (RCQM), Rice University, Houston, 77005, TX, USA}
\author{Alessandro Bombardi}
\email{alessandro.bombardi@diamond.ac.uk}
\affiliation{Diamond Light Source, Harwell Science and Innovation
Campus, Didcot, Oxfordshire, UK}

\begin{abstract}
Here, we use resonant elastic x-ray scattering (REXS) to investigate the inhomogeneity of the zero-field magnetic spin texture of \doped. By using spatially-resolved REXS, we show that the two magnetic transitions at $T_{N_1}$ = 17 K and $T_{N_2}$ = 14 K originate from two nearly degenerate, yet distinct, orthogonal pairs of $\mathbf{q}$-vectors. The corresponding phases are segregated spatially, such that one fraction of the sample comprises coexisting orthogonal spin-density-wave domains, while another fraction hosts coexisting orthogonal helical domains. Additionally, the helical state forms inversion domains indicating that the inversion symmetry is not broken prior the magnetic transition. Our results suggest that the magnetic state is single-\textbf{q} and revealed a large variation of spin textures across a 0.8 $\times$ 1 mm area of the sample surface. These results demonstrate clear differences between the locally and globally probed magnetic textures, typically assumed to be representative of the system as a whole, highlighting the importance of spatially-resolved probes for accurately describing the magnetic behavior of this class of materials.
\end{abstract}

\maketitle

\section{Introduction}
Centrosymmetric magnetic materials that can host chiral magnetic textures and Skyrmion lattice (Skl) phases in the absence of Dzyaloshinskii–Moriya (DM) exchange interactions \cite{hayami2021, takagi2022, hayami2022} have attracted substantial attention within the condensed matter physics community. The microscopic mechanisms responsible for stabilizing these phases remain a subject of ongoing debate, with proposed origins including frustrated Ruderman–Kittel–Kasuya–Yosida (RKKY) interactions \cite{leonov2015, wang2020}. Prototypical examples are the intermetallic Gd/Eu-based compounds, such as GdPd$_2$Si$_3$ \cite{kurumaji2019}, Gd$_3$Ru$_4$Al$_{12}$ \cite{hirschberger2019}, GdRu$_2$Si$_2$ \cite{yasui2020, wood2023}, and the \series~family of materials \cite{moya2022, moya2023, anu2023, anu2024, neubauer2025}. A central step to uncovering these mechanisms is the accurate determination of the magnetic texture. This is often addressed using resonant elastic x-ray scattering \cite{lovesey1996} and neutron scattering techniques \cite{lovesey2015}. In REXS, for instance, the spin texture is typically resolved only within relatively small portions of the sample, with characteristic length scales from a few tens up to a few hundred micrometers. Neutron scattering, on the other hand, generally provides a signal that is averaged over many crystallographic and magnetic domains. In both methods, essential information about the spatial arrangement of domains and the mesoscale structure remains inaccessible, and the inferred global magnetic texture is frequently assumed as representative of the entire system. However, if the system is intrinsically inhomogeneous—for instance, due to quenched disorder arising from chemical doping and static strain fields—this can produce discrepancies between the local and global magnetic textures \cite{dagotto2005}. In such cases, spatially resolved probes become crucial for investigating these kinds of systems.

In this manuscript, we use spatially-resolved REXS to investigate the magnetic textures and domains in \doped. This system crystallizes in the tetragonal \textit{I}4/\textit{mmm} space group with reported lattice parameters $a$ = $b$ = 4.3335(10) $\mathrm{\AA}$, and $c$ = 10.8715(5) $\mathrm{\AA}$ at 120 K \cite{littlehales2024}. The Eu$^{2+}$ ions form square-net layers within the \textit{ab} plane, which are stacked along the \textit{c} axis \cite{stavinoha2018, moya2023}. The magnetic interactions are anticipated to be mediated via the RKKY mechanism, giving rise to long-range magnetic order at low temperatures \cite{littlehales2024}. The Hall resistivity evolves smoothly as a function of  $H//c$ applied magnetic field, and a weak topological Hall effect (THE) has been proposed for this composition \cite{moya2023}, questioning the formation of non-coplanar spin textures. \doped~is located on the Ga-rich side of the \series~phase diagram and, in contrast to the compounds with $0 \leq x \leq 0.5$ \cite{stavinoha2018}, it does not display a charge-density wave (CDW) transition prior to the onset of magnetic phases. This makes this region of the phase diagram particularly suitable for investigating the magnetic behavior without the influence of a precursory symmetry reduction.


Here, we demonstrate that the magnetic texture can vary substantially depending on the specific region of the sample that is probed. In one region, the magnetic order is dominated by orthogonal spin-density waves, whereas another region comprises orthogonal domains of helical states. This observation suggests that the magnetic texture is single-\textbf{q} multi-domain. The phase segregation is found to correlate with the magnetic phase transitions at $T_{N_1}$ and $T_{N_2}$. Furthermore, the helical state forms inversion domains, indicating that in this particular member of the \series~series, the inversion symmetry is not broken prior the magnetic transition. We discuss possible microscopic origins of the multiple spin textures observed in this system in terms of structural or chemical disorder/inhomogeneity and the concomitant local disruption of the RKKY interactions. Our findings provide a clear example of a pronounced spatial inhomogeneity of the magnetic textures in a member of the \series~family and highlight the importance of employing spatially-resolved probes to elucidate the magnetic structure of such systems.

\section{Methods}
\doped\space samples were grown by self-flux method described elsewhere \cite{moya2022, moya2023}. In order to probe the magnetism, Resonant elastic X-ray scattering (REXS) were performed at the I16 beamline at Diamond Light Source \cite{i16_2026} on a as-grown $\sim$2$\times$2 mm$^2$ \doped~sample, with surface normal parallel to the (00L) direction. The x-ray energy was aligned at the Eu \textbf{L}$_{\mathrm{III}}$ edge, around 6.973 keV, and the resonance was confirmed by measuring the fluorescence of the sample through an energy scan. The methodology used for the domain mapping is described in Ref. \cite{anu2024}. The polarization of the incident x-ray beam was controlled using a 400 $\mu$m-tick diamond quarter-wave phase retarder. The azimuthal scans  and the temperature dependence were performed in vertical scattering geometry with an incident beam size of 200 $\mu$m in the horizontal direction and $\sim$ 40 $\mu$m in the vertical direction. The sample was mounted with (00$l$) as the specular direction and the azimuthal reference ($h$00), such that at $\Psi$ = 0 the a axis was parallel to the incident beam. An ARS He-closed cycle refrigerator was used to cool the sample from room temperature to 6.5 K. To analyze the diffracted beam, a Cu (220) crystal oriented close to $\theta\sim$ 45$^{\circ}$ with respect to the incident beam was used. A \textit{QuadMerlin} area detector was used to collect the diffracted intensity after the crystal analyzer.

\section{Results}

\begin{figure}[h!]
\centering
\includegraphics[width=0.5\textwidth]{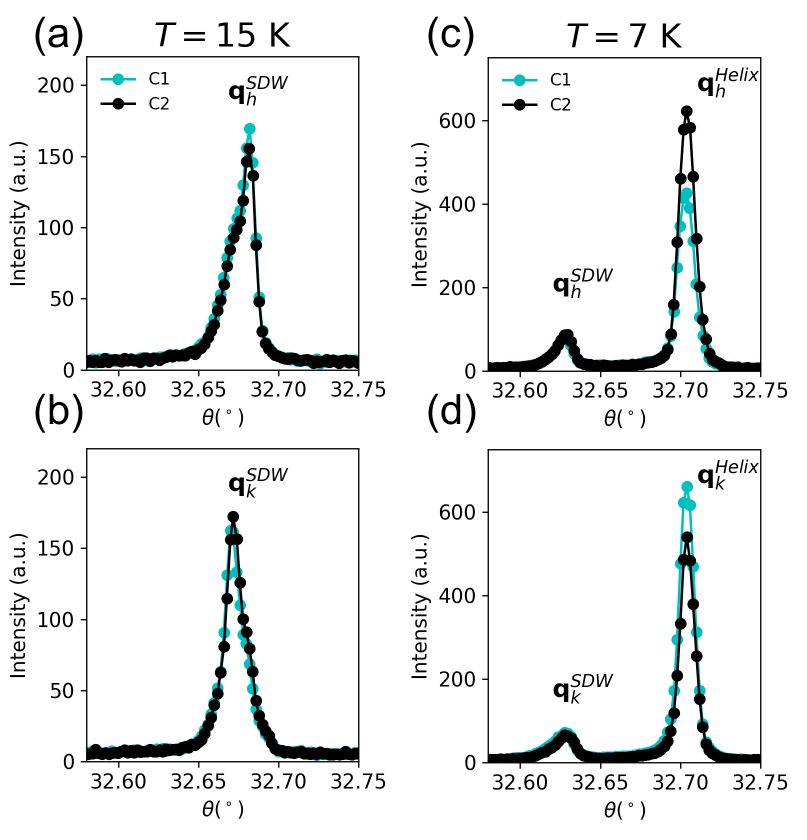}
\caption{Rocking curves around the satellites with opposite circular lights (C1 and C2, respectively) at $T$ = 15 K (b,c) and $T$ = 7 K (d,e) at $\Psi$ = -45$^\circ$. }
\label{fig1}
\end{figure}

Four magnetic satellites of the (0,0,6) Bragg reflection were found. Below $T_{N_1}$ = 17 K, two satellites emerge at \magSDWh~$\approx$ (0.2320(2),0,0) and \magSDWk~$\approx$ (0,0.2319(5),0), followed by the appearance of \magHelixh~$\approx$ (0.2363(5),0,0) and \magHelixk~$\approx$ (0,0.2350(2),0) below $T_{N_2}$ = 14 K in good agreement with our magnetometry measurements and the previously reported transition temperatures in \doped. The rocking curves around the sattelites are shown in Fig.~\ref{fig1} at $T$ = 7 K (a,b) and $T$ = 15 K (c,d). The scattered intensity of all satellites was measured with opposite circularly polarized light. The presence of contrast between opposite circular polarizations (C1 and C2) indicates a non-collinear spin texture, and the absence of such contrast indicates a collinear structure or equally populated inversion domains. As one can see, at 15 K and 7 K, \magSDWh and \magSDWk show no contrast, indicating that the spin texture associated to these wave vectors seems to be preserved across $T_{N_2}$.  On the other hand, \magHelixh and ~\magHelixk appear to be associated to non-collinear structures with opposite handedness, as revealed by the opposite contrast between C1 and C2 lights.
\begin{figure*}[t!]
\centering
\includegraphics[width=1\textwidth]{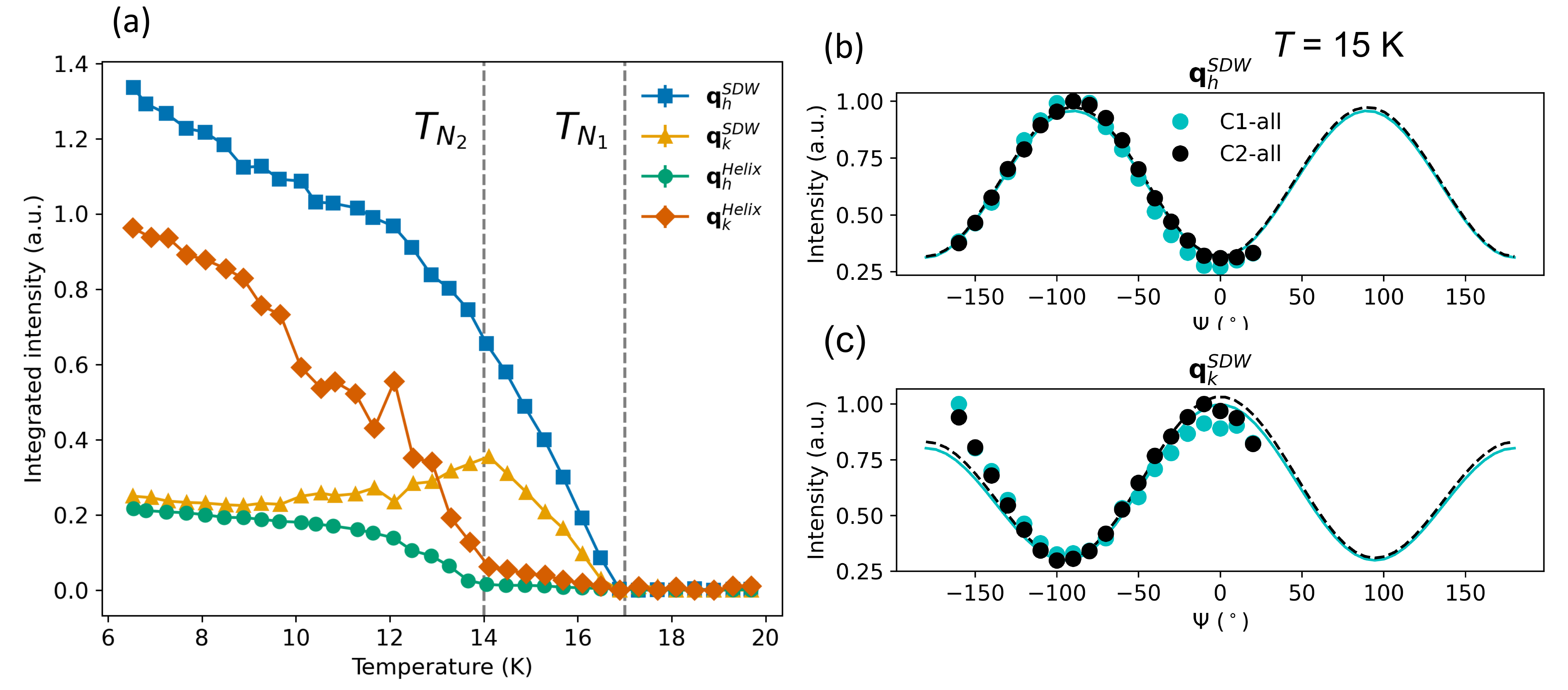}
\caption{Temperature dependency of the integrated intensity of the \magSDWh, \magSDWk, \magHelixh, \magHelixk~satellites measured using a large beam size of 200$\times$40 $\mu$m (a). Azimuthal scans using opposite circularly polarized lights (C1 vs C2) of the \magSDWh (b) and \magSDWk (c) satellites at T = 15 K. The azimuthal reference is ($h$00) [$\Psi$ = 0$^{\circ}$]. The dashed lines represent the fit obtained for the data points.}
\label{fig2}
\end{figure*}
The temperature dependence of the integrated intensity ($I$ vs $T$) of the satellites are shown in Fig. \ref{fig2}(a). It is noticeable the differences in the $I$\ vs $T$ curves for the different satellites. For example, \magSDWh~monotonically increases as the temperature is lowered and it is not significantly affected below $T_{N_2}$. Conversely, \magSDWk~suffers a sudden drop on the onset of \magHelixk. To understand these differences, we measured the temperature dependence at different regions of the sample using a smaller beam size (100$\times$ 20 $\mu$m), revealing a significant variation of the $I\times T$ curves across the sample. For instance, the reduction of the \magSDWh~ intensity below $T_{N_2}$ was mainly detected in those regions where \magHelixh~develops a similar intensity (see Fig. S1 in the Supplemental Material for details) \cite{SM}. This behavior can be ascribed to preferential domain nucleation, which is strongly influenced by the specific region of the sample that is probed, as will be discussed later in the text.

To elucidate the magnetic structure, a standard methodology is to perform azimuthal scans \cite{lovesey1996, beaurepairen2010}, \textit{i.e.}, to rotate the sample around the scattering vector while recording the scattered intensity as a function of the azimuthal angle. This procedure probes the dependence of the scattering cross-section on the relative orientation among the incident polarization, the magnetic moments, and the crystallographic axes. The resulting intensity modulation constitutes a characteristic fingerprint of the moment orientation, allowing for a direct comparison of the experimental data with the magnetic configurations permitted by symmetry. Starting from the parent space group \textit{I}4/\textit{mmm} and incommensurate propagation vectors with the form of \textbf{q} = ($\delta$,0,0) or (0,$\delta$,0), three magnetic irreducible representations (irreps) enter into the decomposition of the full magnetic representation: mSM2, mSM3, and mSM4 \cite{anu2023}. By using a single irrep or using all real (all imaginary) combinations of its basis vectors one can obtain collinear structures that correspond to spin-density-waves. In contrast, non-collinear spin configurations arise when two different irreducible representations are combined, one contributing a real and the other an imaginary basis vector, leading to helical or cycloidal spin structures. In both cases, for collinear or non-collinear solutions, the orientation of the magnetic moments is fixed by the directions of the underlying basis vectors. 

The azimuthal scans for \magSDWh~and \magSDWk~performed at $T_{N_2}<$ $T$ = 15 K $<T_{N_1}$ with C1 and C2 lights are shown in Fig. \ref{fig2}(b,c). First, only a negligible contrast is observed for the whole azimuthal range measured, indicating that the spin textures related to these \textbf{q}-vectors are collinear. The small differences observed between the C1 and C2 curves are attributed to slightly different scale factors of the Fourier components obtained for the individual fits (see Table I in the supplemental material) \cite{SM}. For \magSDWh, the maximum (minimum) occurs at $\Psi$ = -90$^{\circ}$ ($\Psi$ = 0$^{\circ}$) indicating that the magnetic moments are transverse to the direction of \magSDWh. The same conclusion is obtained for \magSDWk. The magnetic structure solution transforms by a single irrep, consisting of spin density waves with moments $\mathrm{m}//b$ for \magSDWh~and $\mathrm{m}//a$ for \magSDWk. The magnetic moments oriented perpendicular to the propagation vector are similar to those observed in EuGa$_2$Al$_2$ \cite{anu2023}. The results of the fittings are available in the supplemental material \cite{SM}. 
\begin{figure*}[t!]
\centering
\includegraphics[width=1\textwidth]{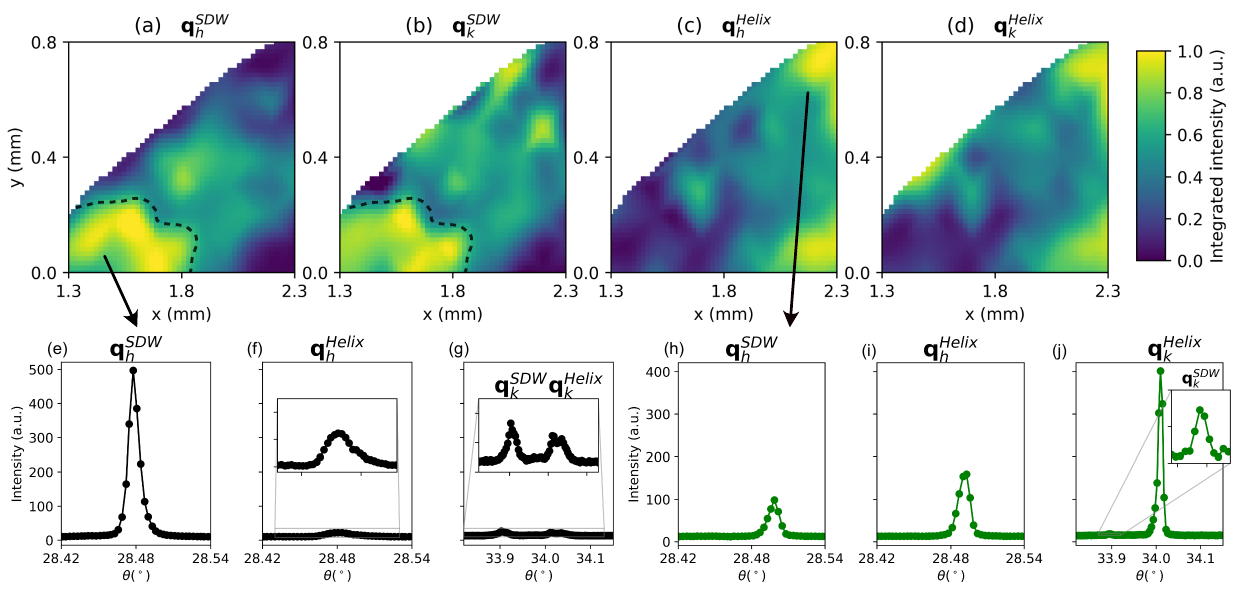}
\caption{Intensity maps showing the spatial variation of the \magSDWh~(a) \magSDWk~(b) \magHelixh~(c) \magHelixk~(d) satellites at 6.5 K. Rocking curves of the satellites at selected regions of the intensity maps indicated by the black arrows (e-j). The peaks display the satellite label on top. All panels share the same intensity scale and the insets have individual zoomed scales to show the presence of the low intensity peaks.}
\label{fig3}
\end{figure*}

Below $T_{N_2}$, however, the presence of multiple domains associated to possibly different spin textures in close proximity in reciprocal space hinders an unambiguous assignment of the magnetic structure to each satellite, since the scattered intensities associated with two distinct propagation vectors, corresponding to different spin textures are often superimposed in the azimuthal scans (See Fig. S2 in the supplemental material) \cite{SM}. In our measurements below $T_{N_2}$ [Fig. \ref{fig1}(d,e)] the absence of contrast at $\Psi$ = -45$^{\circ}$ observed for \magSDWh~and \magSDWk is preserved, indicating that the spin texture of these satellites is likely the same as obtained at $T$ = 15 K. Conversely, the contrast between opposite circular lights at $\Psi$ = -45$^{\circ}$ detected for \magHelixh~and \magHelixk~could, in principle, result from either a helical or a cycloidal spin texture; however, as will be shown later in the text,  C1 vs C2 contrast maps carried out at an azimuthal angle of $\Psi\sim$ -90$^{\circ}$ demonstrated a null (maximal) contrast for \magHelixh~(\magHelixk). For this particular \textbf{q}-vector and azimuth, a helical structure is the only spin configuration that can yield zero (maximal) contrast (See section III of the supplemental material for details) \cite{SM}. The SDW and helical states found with propagation vectors along the same direction below $T_{N_2}$ are consistent with the results obtained for \doped~in Ref.~\cite{littlehales2024}. However, as we will show in the following sections, some regions of the sample are dominated by orthogonal SDW domains, while others by orthogonal helical domains, which can readily result in an incorrect determination of the magnetic spin texture if one focus in specific regions of the sample.
\begin{figure*}[t!]
\centering
\includegraphics[width=1\textwidth]{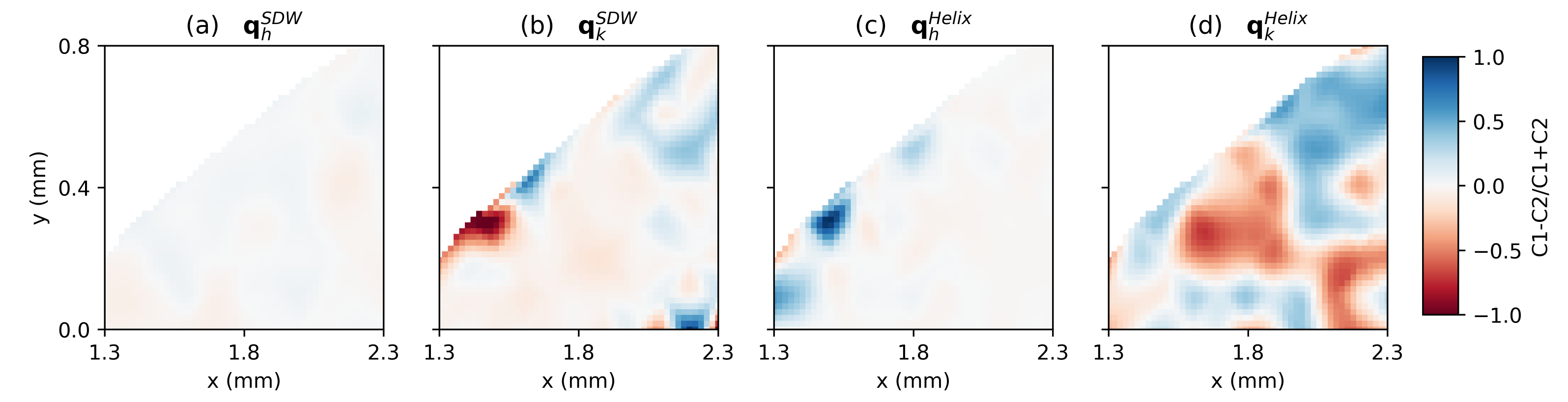}
\caption{Intensity contrast maps from opposite circular lights C1 and C2 for each sattelite \magSDWh~(e) \magHelixk~(f) \magHelixk~(g) \magHelixk~(h) of the (0,0,6) reflection at 6.5 K.}
\label{fig4}
\end{figure*}

The first fundamental question we aim to answer is whether the magnetic state forms a single-\textbf{q} or a double-\textbf{q} structure. In a double-\textbf{q} structure the free energy terms of the \textbf{q}-vectors are coupled creating a secondary lattice modulations of the type \magSDWh~+~\magHelixk~\cite{wood2023, khanh2020, matsumara2024}. Although these type of modulations were not observed here or in previous reports~\cite{littlehales2024}, the absence of such reflections does not serve as a definitive proof that the system is single-\textbf{q}. In addition to the latter, the ratio of the integrated intensity between \magHelixk~and \magSDWh~should be constant across the region wherein the multi-\textbf{q} domain is stabilized \cite{paddison2021}. To address this question and to investigate the magnetic domain texture of \doped, the X-ray beam was reduced to a 100 $\times$ 20 $\mu$m spot, and the sample was scanned over a 1 $\times$ 0.8 mm$^2$ area while the four satellites around the (0,0,6) nuclear reflection were recorded. The spatial variation of integrated intensities obtained from a Gaussian profile fit of all four satellites are shown in Fig.~\ref{fig3}(a–d).  

The \magSDWh~intensity is mainly localized in the lower left corner, corresponding to the brighter area delineated by the black dashed line. The \magSDWk~satellite [Fig.~\ref{fig3}(b)], which is orthogonal to \magSDWh, appears to predominantly occupy the same region as \magSDWh. Although the overall intensity of \magSDWk~ is lower compared to \magSDWh, the shape of the intensity distribution is remarkably similar, as detailed by the dashed line Fig.~\ref{fig3}(b). Conversely, \magHelixh~is concentrated on the upper right corner of the sample as well as \magHelixk, also showing similar patterns of the intensity distribution. Moreover, the region which is occupied by \magSDWh~shows a much lower intensity distribution of \magHelixh~and \magHelixk. This is explicitly illustrated by inspecting the satellites intensity at selected regions of the sample as shown in Fig. \ref{fig3}(e-j); The regions are indicated by the black arrows. At the lower left region, there is an over one order of magnitude difference in the intensity  between \magSDWh~and the other satellites [Fig. \ref{fig3}(e-g)]. On the extreme opposite side, \magHelixh~and \magHelixk~dominates, with larger intensity compared to \magSDWh and \magSDWk. Although all four satellites are always present in the sample, the observation of such large variation in their intensity ratios combined with the absence of \magSDWh~+~\magHelixk~type reflections suggests that the magnetic state in this material is single-\textbf{q}. Despite the spatial variation in scattering intensity across the sample, the primary phase segregation is delineated by the two pairs of satellites \magSDWh/\magSDWk~and \magHelixh/\magHelixk, corroborating the temperature dependence that shows that \magSDWh/\magSDWk~ is associated with $T_{N_1}$, whereas \magHelixh/\magHelixk emerges at $T_{N_2}$. This observation indicates a competition between these two magnetic states. The competition between these two states is further supported by measurements upon warming the sample to 15 K, \textit{i.e.}, $T_{N_2}$$<T<$$T_{N_1}$, where only SDW are observed. At this temperature, the intensity of the \magSDWh~satellite extends over the region previously occupied by \magHelixh~and \magHelixk~(See Fig. S3 in the supplemental material) \cite{SM}.

To understand if the spin texture is consistently maintained, the sample was rastered using two opposite circular lights and the contrast maps, \textit{i.e.}, the intensity difference between C1 and C2, for each satellite are shown in Fig. \ref{fig4}(e-h). The satellites \magSDWh, \magSDWk, and \magHelixk~show null or negligible contrast across the entire sample. We note that the absence of contrast in \magHelixk~is due to the fact that maps were performed at $\Psi\sim$ -90$^\circ$, where the contrast for this \textbf{q}-vector is expected to be suppressed. Conversely, the \magHelixk~satellite shows a large variation of the C1 vs C2 contrast across the sample, indicating the formation of inversion domains of the helical state. This is in contrast to what was observed in the parent compound \parent, wherein the chirality of the helical state is preserved across a large area of the sample~\cite{anu2024}.  

\section{Discussion}
Our results demonstrate that \magSDWh~/\magSDWk~and \magHelixh/\magHelixk~belong to two distinct magnetic phases associated with $T_{N_1}$ and $T_{N_2}$, respectively. These order parameters exhibit different temperature dependencies and a clear signature of competition, in which the volume fraction of the domains depend on the region of the sample that is probed. Our results suggests a single-$\mathbf{q}$ multi-domain configuration for \doped, as evidenced by the spatial segregation and intensity variation of the satellites across the sample. The most prominent phase segregation occurs between the pairs of satellites \magSDWh/\magHelixk~and \magHelixk/\magHelixk. Despite corresponding to orthogonal spin-density-wave domains, \magSDWh~and \magHelixk~are found to occupy predominantly the same regions of the sample. This observation suggests that the characteristic domain size associated with these orders are smaller than the beam footprint, and that the areas in which they appear to coexist are likely composed of a interleaving of such domains. The larger intensity of \magSDWh~relative to \magSDWk~indicates that, although both are concentrated in the same spatial regions, \magSDWh~domains have a larger volume fraction. 
A similar behavior was observed for the \magHelixh~and \magHelixk~ satellites, which occupy the same region the sample. In this case,  \magHelixk, shows a larger volume fraction than \magHelixh. 
In addition, we observed the presence of inversion domains in the helical state associated to \magHelixk, \textit{i.e.}, regions of the sample exhibiting helices with opposite handedness. This observation indicates that the system does not select a unique chiral state, providing evidence that inversion symmetry is not broken until the magnetic phase transition. This is especially significant, as it is probably connected to the lack of a CDW state or a monoclinic distortion in \doped, which has been suggested to break the inversion symmetry in \parent~and EuGa$_2$Al$_2$ \cite{anu2023,anu2024, kotla2025}. The presence of such symmetry breaking preceding the magnetic state would allow for the formation of a single-chiral state and DM-type interactions, responsible for the formation of Skyrmion phases and other non-coplanar spin textures. 


In rare-earth based intermetallics, the main driving mechanism to form the complex magnetic states is based on RKKY interactions \cite{hayami2021, takagi2022}. The RKKY interaction has an oscillatory long-range nature allowing for ferromagnetic or antiferromagnetic coupling depending on the distance between the localized magnetic moments in the real space. The coupling strength typically decay in real space as $\sim$ 1/r$^d$, where $d$ is the dimensionality of the system. However, the presence of disorder and local defects can effectively suppress its long-range nature, modifying the magnitude and the decay of the exchange interactions \cite{sobota2007}.  Unlike the Ga-deficient members of the \series~series, where the crystal symmetry is reduced from tetragonal to orthorhombic across the CDW transition \cite{anu2023}, \doped~shows no indication of a symmetry reduction prior to the magnetic transitions. Consequently, the four-fold symmetry is expected to remain intact in the magnetic states. Nevertheless, we observe \textbf{q}-vectors associated with distinct spin textures, accompanied by strong spatial variations in their domain populations across the sample. We propose that local variations in defects and disorder likely influence the RKKY interactions, giving rise to a set of nearly degenerate competing magnetic ground states. Indeed, degenerate Fermi-surface-nesting \textbf{q}-vectors and the corresponding RKKY interactions have been suggested to play a crucial role in stabilizing the magnetic phases in \parent~\cite{miao2024, arai2026}. 
Throughout the \series~series, substituting Ga with Al leads to dramatic changes in magnetic behavior, driven by the nonmonotonic evolution of the $a$ lattice parameter and the Eu–Eu bond lengths \cite{stavinoha2018}, consisting in a deviation from Vegard's law \cite{denton1991}.  
Accordingly, local variations in Ga and Al concentrations would also locally alter the Eu–Eu bond distances, thereby modifying the nearest-neighbor RKKY interactions. In this context, local probes such as micro-EXAFS could shed light on spatial variations in the Eu local environment and their link to the emergence of different spin textures. Another scenario involves vacancies and/or crystal defects acting as nonmagnetic disorder, which can change the conduction electron wave functions and generate regions of strong localization, thereby transforming the RKKY interaction from a power-law to an exponential decay \cite{sobota2007}.

Finally, we emphasize that our results indicate that conventional approaches to determining the magnetic texture of such materials (whether based on measurements of small, localized regions of the sample or on extensive spatial averaging) can lead to an incorrect identification of the true magnetic state.


\section{Summary} 
Our results show that the Ga-rich compound \doped~stabilizes a complex magnetic texture. The two magnetic transitions seen in bulk measurements arise from two nearly degenerate, orthogonal sets of $\mathbf{q}$-vectors. Spatially resolved measurements reveal phase segregation between these competing states: one sample fraction hosts coexisting orthogonal spin-density-wave domains, while another contains coexisting orthogonal helical and spin-density-wave domains. We also find that the inversion symmetry is preserved in \doped,  which places the Ga-rich side of the \series~phase diagram as good prototypes for understanding whether these systems can host Skyrmion phases and other topological textures in the absence of structural changes that lead to the break of the inversion symmetry preceding the magnetic phases. This intricate spatial variation of the magnetic structure motivates further spatially resolved studies of the \series~series to clarify how Ga substitution affects the domain configuration of the magnetic spin textures. Our work also highlights the strengths of spatially-resolved REXS as probe for systems with substantial inhomogeneity and spatial variation in the spin textures. It underscores the importance of spatially resolved probes for disentangling competing magnetic textures in intermetallic topological magnets.

\begin{acknowledgments}
We acknowledge Diamond Light Source for time on beamline I16 under Proposals No. NT42273-1, MM41580-1, MM41580-2 and MM43885-1. Work at Rice University was supported by the Robert A. Welch Foundation under Grant No. C-2114.
\end{acknowledgments}
\bibliography{lib}
\end{document}